\documentclass[12pt]{iopart}

\usepackage{graphicx}
\usepackage{xcolor}
\usepackage{soul}
\usepackage[compress]{cite}
\usepackage{siunitx}

\begin{document}

\title[A cryogenic buffer gas beam of BaOH molecules using water vapor seeded neon gas]{A cold beam of BaOH molecules using a water-vapour seeded neon gas}

\author{Ties H. Fikkers$^{1,2}$}
\author{Nithesh Balasubramanian$^{1,2}$}
\author{Joost W.F. van Hofslot$^{1,2}$}
\author{Maarten C. Mooij$^{2,3}$}
\author{Hendrick L. Bethlem$^{1,3}$}
\author{Steven Hoekstra$^{1,2}$}

\address{$^1$ Van Swinderen Institute for Particle Physics and Gravity, University of Groningen, The Netherlands}
\address{$^2$ Nikhef, National Institute for Subatomic Physics, Amsterdam, The Netherlands}
\address{$^3$ Department of Physics and Astronomy, and LaserLaB, Vrije Universiteit Amsterdam, The Netherlands}

\ead{s.hoekstra@rug.nl}

\begin{abstract}
In this paper we report on the production and characterization of a cold beam of BaOH molecules using a cryogenic buffer-gas beam source. BaOH is a highly suitable molecule for studies of the violation of fundamental symmetries, such as the search for the electron's electric dipole moment. BaOH molecules are synthesised inside the cold source through laser ablation of a barium metal target while water vapor is seeded into the neon buffer gas. The BaOH flux is significantly enhanced ($\sim$11 times) when laser-exciting the barium atoms inside the buffer-gas cell on the $^1\mathrm S_0 - ^3\mathrm P_1$ transition. A similar enhancement has been reported for other alkaline-earth(-like) monohydroxides. For typical source conditions, the molecular beam has an average velocity of $\approx180$ m/s and an intensity of $\sim 10^{9}$ molecules s$^{-1}$ in $N=1$, which is comparable to that of cryogenic BaF beams.
\end{abstract}

\maketitle

\section{Introduction}

Polyatomic molecules have recently gained attention for their applications to precision tests of fundamental physics~\cite{Roussy.Cornell.2023, Andreev.ACMECollaboration.2018, Safronova.Clark.2018, Kozyryev.Doyle.2021, Yu.Hutzler.2021}. The increased complexity of their internal structure compared to that of atoms and diatomic molecules opens up a wide range of opportunities across various fields, including quantum simulation, quantum computing, measurements of fundamental constants, advanced quantum control, and cooling techniques~\cite{Wall.Carr.2015, Yu.Doyle.2019, Albert.Preskill.2020, Hutzler.Hutzler.2020, Anderegg.Hutzler.2023, Augenbraun.Doyle.2023}. Similar to heavy polar diatomic molecules like BaF, the presence of a strong internal electric field in polyatomic molecules results in a heightened sensitivity to the electron's permanent electric dipole moment (\textit{e}EDM)~\cite{Bause.Hoekstra.2025, Chamorro.Pasteka.2022}.

Linear triatomic molecules possess a `bending' vibrational mode that gives rise to low lying closely spaced states of opposite parity. These close lying states mean the molecule is easily polarized in a low electric field and can act as co-magnetometer states. This provides us with a powerful tool to suppress systematic sources of uncertainty in precision experiments. Laser cooling is an important property, because it enables the accumulation and trapping of molecular samples, boosting the sensitivity of precision measurements. No diatomic species has been found that offer both these properties. It is, however, relatively straightforward to find polyatomic molecules that offer both the possibility of laser cooling and internal co-magnetometer states.  In recent years, linear alkaline-earth(-like) metal monohydroxide species similar to BaOH such as CaOH~\cite{Takahashi.Miyamoto.2022}, SrOH~\cite{Kozyryev.2017,Kozyryev.Doyle.2021} and YbOH~\cite{Jadbabaie.Hutzler.2020} have been experimentally investigated. The production of intense and slow beams of these molecules in the desired quantum state is the crucial starting point of all ultracold molecule experiments. The knowledge gained from the optimised production of such molecules is also directly applicable to similar molecules containing radioactive nuclei relevant for precision tests of fundamental physics, like RaOH~\cite{Isaev.Eliav.2017}.

In this paper, we present our results on the production of a cryogenic buffer gas beam of BaOH as the first step towards an \textit{e}EDM experiment using optically trapped BaOH in the fundamental vibrational `bending' mode~\cite{Bause.Hoekstra.2025}.

\section{Production of BaOH}

Barium monohydroxide was already observed in flames containing alkaline earth salts by Herschel in 1823 \cite{Herschel.1823} and was first definitively assigned by James and Sudgen in 1955 \cite{James.Sugden.1955}. Subsequent successful production of the $^2\Sigma$ ground state radical was achieved in Broida-type ovens, where barium metal is vaporised in a crucible, entrained in a carrier gas flow (typically argon) and then made to react with a reactant gas (such as H$_2$O, D$_2$O, H$_2$O$_2$ and CH$_3$OH) in a low pressure environment \cite{Kinsey-Nielsen.Bernath.1986,Fletcher.Ziurys.1995, Wang.Bernath.2008}. Supersonic beams of BaOH have also been produced, by laser ablating a rotating barium rod in a supersonic expansion of argon carrier gas seeded with room-temperature methanol~\cite{Frey.Steimle.2011}.

However, for the purposes of laser cooling and subsequent precision experiments, the large forward velocity of a supersonic beam severely limits the available interaction time. Producing these molecules in a cryogenic buffer gas beam (CBGB) source \cite{Hutzler.Doyle.2012} offers an efficient solution for generating molecular beams in low-rotational states with a higher intensity at a lower forward velocity. In such a source, radicals are produced by the reaction of an ablated precursor with a reactant gas, inside a cooled cell filled with an inert buffer gas. The buffer gas internally cools the produced molecules while also entraining them as they flow out through an aperture, producing a beam with a forward velocity on the order of 150 -- 180~m/s. 

The linear triatomic molecules CaOH, SrOH and YbOH have been produced in CBGB sources through ablation of salt precursor targets~\cite{Takahashi.Miyamoto.2022, Kozyryev.2017, Jadbabaie.Hutzler.2020}. The production using salt targets has been optimised by following a very specific mixture and recipe \cite{Hutzler.Doyle.2011, Barry.DeMille.2011}. Salt targets are characterised by low thermal conductivity and high heat of vaporisation ($\sim 0.9$ W/(\(\text{m}\cdot\text{K}\)) and $\sim10^4$ kJ/mol respectively for Ba(OH)$_2$\cite{Wang.Zhao.2019})). In the case of pulsed nanosecond laser ablation, higher laser fluences are required in order to exceed the ablation threshold, compared to pure metallic targets that have high thermal conductivity and low heat of vaporisation \cite{Chichkov.Tunnermann.1996}. The disintegration of the salt target on the measurement timescale of a week is suspected to arise from its high fluence threshold combined with its fragility. Furthermore, successive ablation of brittle materials cause exacerbating inhomogeneities on the surface, which underlie large fluctuations in the yield \cite{Korner.Bergmann.1996}. These issues can be overcome or compensated by having a chemically advantageous target \cite{West.2017} or by using direct millisecond pulse laser heating instead of nanosecond ablation pulses \cite{Winnicki.Doyle.2024}. However, producing targets that last for an appreciable amount of time and produce a stable yield of molecules can be challenging \cite{Jadbabaie.Hutzler.2020, Hutzler.Doyle.2012, Barry.DeMille.2011, Skoff.Hinds.2009}. An alternative approach is the use of a metal target, which has been used successfully in the production of diatomic molecules in both supersonic beams~\cite{Tarbutt.Ezhov.2002, Aggarwal.Zapara.2018} and CGBG beams \cite{Mooij.Yin.2024, Wright.Truppe.2023}. By comparison, elemental barium possesses larger thermal conductivity and lower heat of vaporization (18~W/(m~$\cdot$~K) -- 40~W/(m~$\cdot$ K) \cite{Ho.Liley.1972} and 140~kJ/mol respectively \cite{Zhang.Yang.2011}). Consequently, the ablation power needed to ablate metal targets is considerable lower when compared to salt targets, which leads to a larger and more stable molecular flux. The source can run for a longer time before maintenance is required. The lower thermal load on the cell due to reduced ablation power also favours low beam velocities and low rotational temperatures. However, metal targets do come at the cost of having to add OH groups in the form of a second target or a gaseous precursor such as water vapor into a cryogenic environment. 

In experiments with Broida ovens, it was found that yields of CaOH and SrOH increased significantly upon exciting the calcium or strontium atoms from the $^1\mathrm{S}_0$ ground state to the metastable $^3\mathrm{P}_1$ state ~\cite{Brazier.Bernath.1986} . It was argued that the additional energy in the atoms makes the reaction exothermic. We point out that due to the spin-forbidden nature of the transition, the $^3\mathrm{P}_1$ states have a relatively long lifetime ($\geq 1$ ~\unit{\micro s}), decaying either back to the $^1\mathrm{S}_0$ state or to $^3\mathrm{D}_1$ and $^3\mathrm{D}_2$ states \cite{Scielzo.Potterveld.2006}, thereby  providing sufficient time for the atoms in the excited state to form monohydroxides inside the cell. This yield enhancement was later also observed for both BaOH and BaOD \cite{Fernando.Bernath.1990}. To understand the underlying thermochemistry behind the production reaction, Davis et al. \cite{Davis.Mestdagh.1993} studied the effect of BaOH yield by exciting the barium atoms  on the $^1\mathrm{S}_0-^1\mathrm{P}_1$ transition at 553~nm (see Figure \ref{fig:baoh-spectrum-probe} (b)), which also gave an increased yield.
More recently, work was done to enhance the production of YbOH in the cryogenic buffer gas environment by exciting atomic ytterbium to the $^3\mathrm{P}_1$ state~\cite{Jadbabaie.Hutzler.2020}. In this article we investigate the enhanced production of BaOH molecules via the laser excitation of atomic barium. We use the $^3\mathrm{P}_1$ state, since it is accessible with our laser systems.

We have compared the production of BaOH from salt and metal targets, used both methanol and water as reactant gas, and explored the enhancement of production of BaOH by exciting the electronic state of the barium atom. Once settled on the optimal production scheme, we have characterized the forward velocity and intensity of the molecular beam as a function of the carrier gas flow rate and the reactant gas concentration.

\section{Methods}
In this section, we provide details of our experimental setup and methods.

\subsection*{Cryogenic cell design}

\begin{figure}
    \centering
    \vspace{0.5cm}
    \includegraphics[width=0.8\linewidth]
    {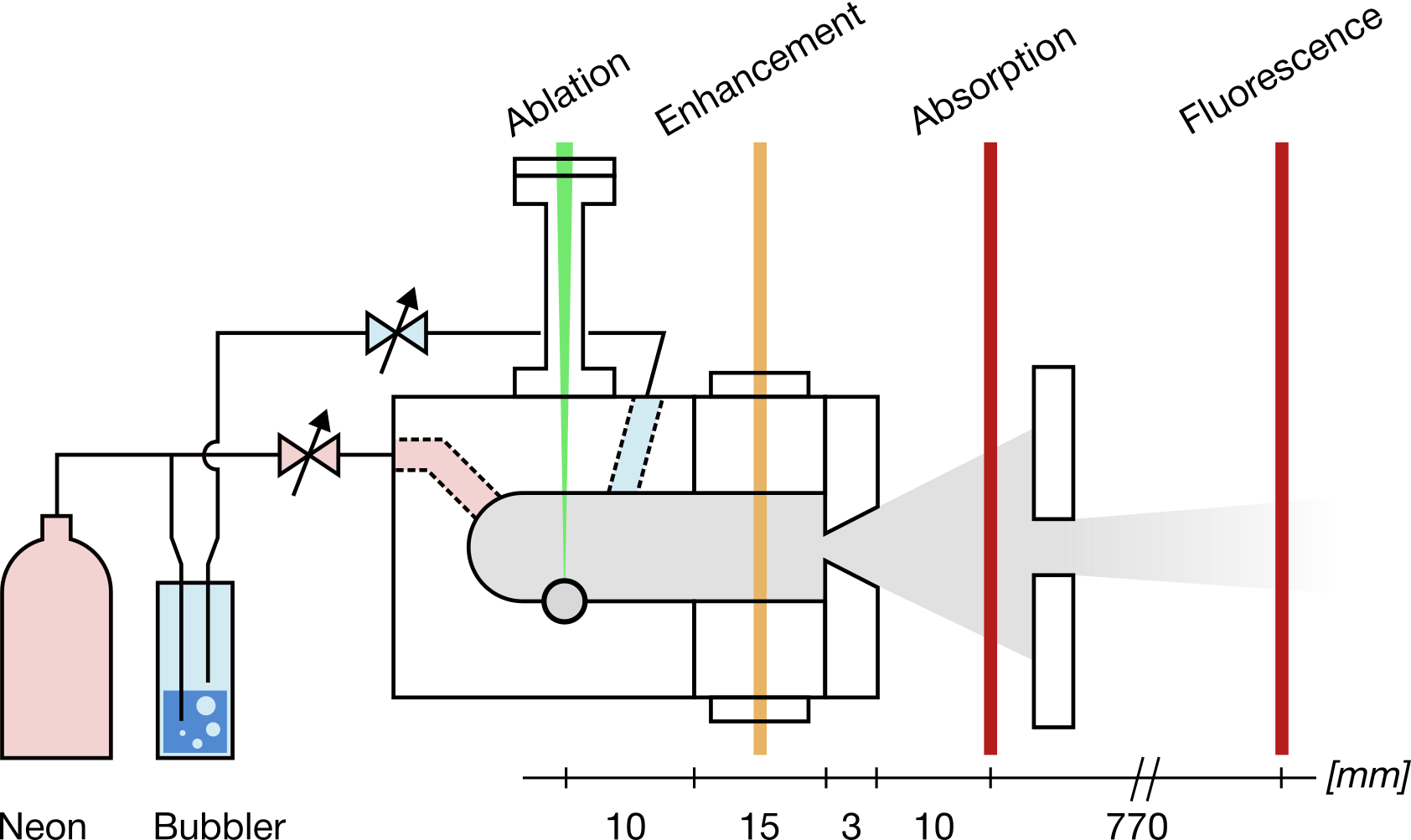}
    \vspace{0.5cm}
    \caption{%
    Schematic of the buffer gas source and the absorption and fluorescence detection zones. BaOH molecules are produced by laser ablating a rotating barium target in the presence of cold neon buffer gas and water-vapour seeded neon gas. The neon line is cooled by thermal contact to the cryocooled surfaces, the reactant mixture comes from a bubbler filled with water through a heated fill line. Molecular yield is enhanced by sending in laser light through an optical access window orthogonal to the cell.}
    \label{fig:experimental-setup}
\end{figure}

The cryogenic buffer gas source used is the same as used in previous work \cite{Mooij.Willmann.2024, Mooij.Yin.2024}, which is based on prior designs \cite{Esajas.2021, Truppe.Tarbutt.2018}. A cell extension with optical access was added to facilitate the laser excitation of barium. Key parts of its design are schematically shown in Figure~\ref{fig:experimental-setup}. Aluminum and copper radiation shields are mounted to the first ($\sim$40~K) and second ($\sim$4~K) stages, respectively, of a cryo-cooler (Sumitomo Heavy Industries, RP-082B2S). The buffer gas cell is mounted to the second stage through an adjustable thermal resistance. The cell temperature is regulated to about 20 K using a heater cartridge. Neon buffer gas, precooled to 20 K, is introduced through the back of the cell at a typical flow rate of 20 sccm (Bronckhorst FG-201CV). A reactant gas can be supplied through a heated fill line, flowing into the cell through a thermally insulated separate entrance. The cell itself is made up of a 30 mm square copper block with a 10 mm diameter cylindrical inner bore. The cell has a 15 mm long extension which provides optical access to the cell bore through sapphire glass windows. This can be used for in-cell spectroscopy and for excitation of barium atoms. A cylindrical barium metal target is inserted into the cell from below and is rotated and translated using a motor. Light from a pulsed Nd:YAG ablation laser (Spectra-Physics Quanta-Ray GCR-12S, 10 ns pulses of 532 nm at 10 Hz) enters from the side through a sapphire window offset from the cell by a tube segment. The molecular beam leaves the cell through a 4.5 mm diameter aperture.

\subsection*{Molecule synthesis}
We have tested Ba(OH)$_2$, Ba + H$_2$O, Ba + CH$_3$OH as methods for producing BaOH. First we tried ablating a salt Ba(OH)$_2$ target. The targets were made using Ba(OH)$_2$ power mixed with epoxy resin. This mixture was then poured into a mold and cured in a vacuum chamber at 200 mbar. BaOH signal was observed but its yield depended greatly on the position of the ablation spot on the target.

Due to the inconsistent yield from salt targets, the focus was shifted to a production method that pairs a metallic barium target with a reactive gas. Methanol was tried as a reactant gas; however, no BaOH was observed using this method hence we turned to H$_2$O, which did give reliable signal. The 273 K reactant gas (in our case a mixture of neon and water) is introduced from the side, with its flow controlled by a gas flow controller, with typical flow rates between 10--20 SCCM. The neon gas is seeded with water by flowing the neon gas through a bubbler filled with distilled water. By controlling the neon head pressure (the pressure above the liquid) it is possible to control the saturation of the neon carrier gas. Preferably the head pressure is close to the vapor pressure of the liquid inside. The water seeded neon gas enters the cell through a temperature controlled fill line usually set to 273 K.

\subsection*{Detection}

\begin{figure}
    \centering
    \vspace{1cm}
    \includegraphics[width=0.8\linewidth]{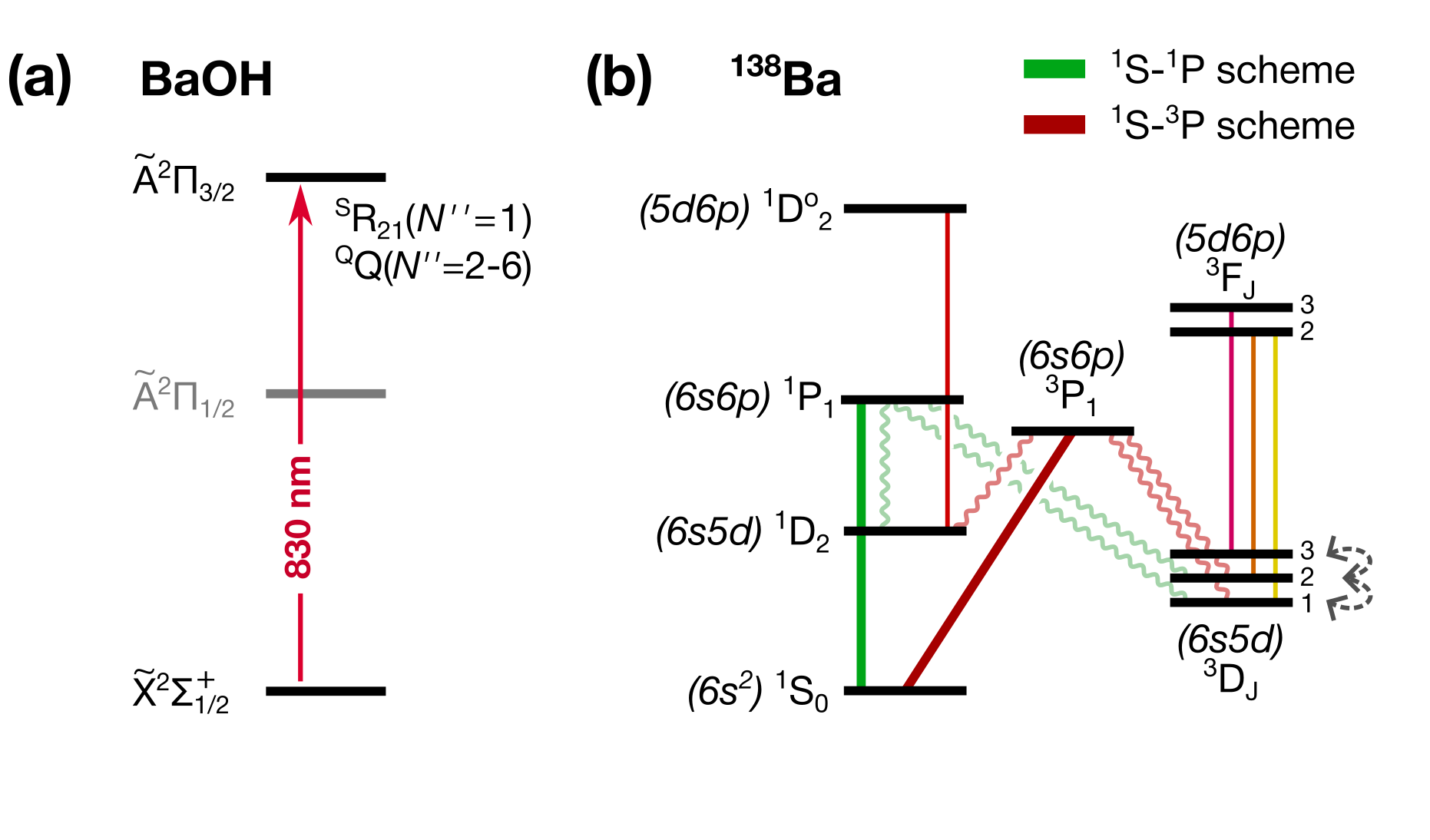}
    \vspace{0.2cm}
    \caption{{\bf (a)} Transitions used to characterise the yield of BaOH, denoted by $^{\Delta N}\Delta J(N_\text{ground})$. The $^SR$ transition was generally used for characterising the molecular yield while the $^QQ$ transitions were used for estimating the rotational temperature of the beam.~{\bf (b)} Relevant energy levels in neutral $^{138}$Ba. In green is the pumping scheme associated with driving the transition $^1\mathrm S_0 - ^1\mathrm P_1$ to enhance the production of BaOH, as reported in \cite{Davis.Mestdagh.1993}. The effect of this pumping scheme is to transfer the $^1\mathrm S$ population into the metastable $^1\mathrm D$ population. In red is the pumping scheme associated with driving the transition $^1\mathrm S_0 - ^3\mathrm P_1$, which is described in the text. Here, the barium population primarily decays to the metastable $^3\mathrm D_2$ state. Also shown are the transitions used to probe the metastable D state populations. The grey dashed lines indicate collisional transfer between the metastable $^3\mathrm D_J$ states~\cite{Ehrlacher.Huennekens.1994}.}
    \label{fig:baoh-spectrum-probe}
\end{figure}

The molecules can be detected using both absorption and laser-induced fluorescence, as shown in Figure~\ref{fig:experimental-setup}. A Coherent 899 Titanium:Sapphire (Ti:S) laser is used to produces the NIR light used for molecular spectroscopy. This is split into a 5~\unit{\micro W} beam for absorption spectroscopy and a 150~\unit{\micro W} beam for laser-induced fluorescence (LIF) spectroscopy. The absorption laser beam passes the molecular beam 10~mm downstream from the cell aperture and is retro-reflected after which the light is collected using an amplified photo diode (Thorlabs PDB210A/M). The light for fluorescence passes through the molecular beam 780~mm downstream from the cell aperture. The main transition used for these LIF measurements is from the $\tilde A ^2 \Pi_{3/2}(000)$ to the $\tilde X^2\Sigma_{1/2}(000)$ level shown in Figure~\ref{fig:baoh-spectrum-probe} (a). The fluorescence light from the molecules is then collected by a lens system into a PMT (9658A EMI Electronics).

Part of the laser light is sent to a wave\-length meter (Atos LM-007) to measure the frequency with an uncertainty of 30 MHz. Further more a beat note between the laser light and light from a frequency comb (Menlo Systems FC1500-250-WG) stabilized against a caesium clock (Microsemi CSIII Model 4301B) leads to an absolute precision to the order of 1 MHz. The frequency of the excitation laser is digitally locked to the frequency comb to ensure stability throughout our measurements.

Likewise, a second Coherent 899 Ti:S produces laser light ($\sim$100~mW, 791~nm) that is used to excite the $^1\mathrm{S}_0 - ^3\mathrm{P}_1$ transition in the barium atoms, as denoted in Figure~\ref{fig:baoh-spectrum-probe} (b). The laser is locked to the transition using the method described above.

\section{Results}

The production of BaOH was observed with both salt and metal targets. We find that the yield is more consistent and reliable using the metal target, therefore the subsequent measurements used a metal target.

\subsection*{Atomic barium excitation}
\begin{figure}[!ht]
    \centering  
    \includegraphics[width=\linewidth]{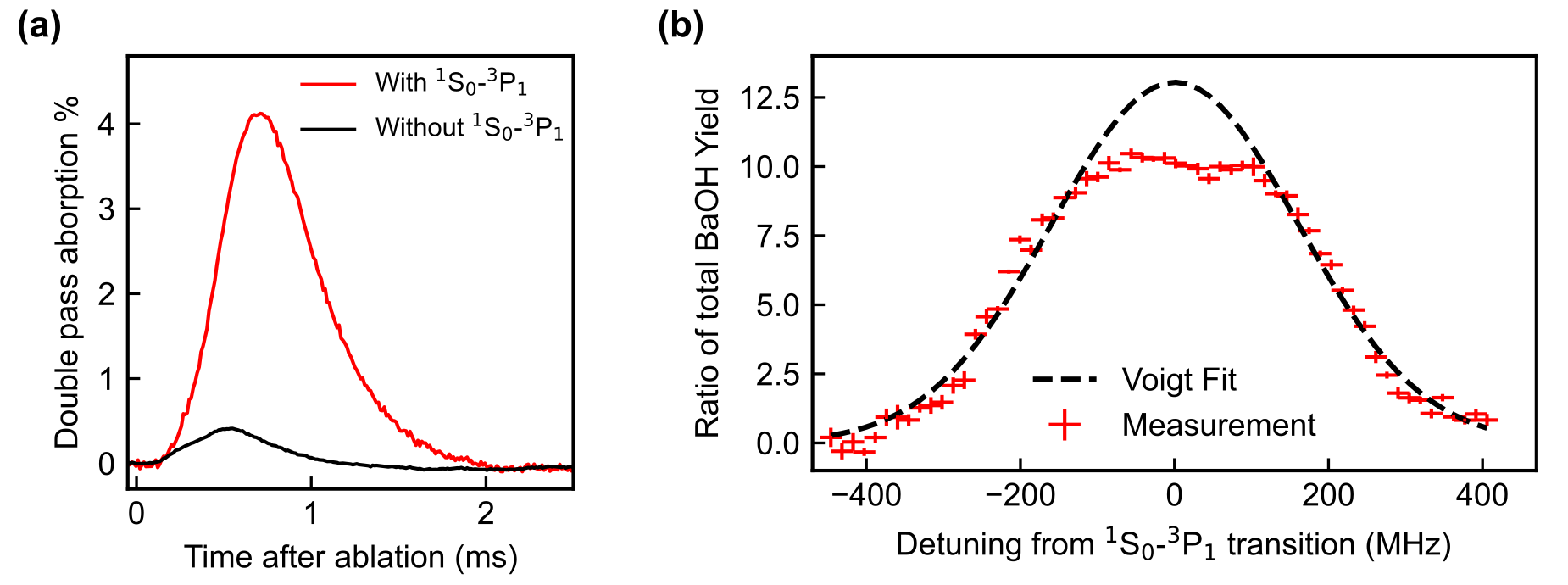}
    \caption{Increased yield of BaOH as a result of the excitation of atomic barium. \textbf{(a)} Time-of-flight curves seen from double pass absorption when absorption laser is locked to the $^SR_{21}(1)$ transition between $\tilde A^2\Pi_{3/2} - \tilde X^2\Sigma_{1/2}$ in the vibrational ground state, with (in red) and without (in black) enhancement laser close to the $^1\mathrm S_0 - ^3\mathrm P_1$ resonance. \textbf{(b)} The enhancement factor (ratio of the OD integrated over the length of the pulse) as a function of the detuning from the barium $^1\mathrm S_0-^3\mathrm P_1$ transition. The enhancement factor saturates to a value around 10 when we are within 200~MHz of the resonance.}
    \label{fig:enhancement}
\end{figure}

\indent Without excitation of atomic barium from the $^1\mathrm{S}_0$ to the $^3\mathrm{P}_1$ level, typically peak double pass absorption percentages of about 0.5--1 \% were observed. After exciting the atomic barium, absorption peak percentages of 4--8 \% were observed, as shown in  Figure~\ref{fig:enhancement} (a). In Figure \ref{fig:enhancement} (b) the yield of enhanced BaOH, normalized against unenhanced BaOH, is plotted against the detuning of the 791~nm laser (100~mW). Based on the measured lifetime of the $^3\mathrm{P}_1$ state the natural linewidth of this transition is expected to be about 743(8)~kHz \cite{Scielzo.Potterveld.2006}. The width of the peak in Figure \ref{fig:enhancement} (b) is dominated by thermal Doppler broadening and power broadening.

\begin{figure}[!ht]
    \centering
    \vspace{0.5cm}
    \includegraphics{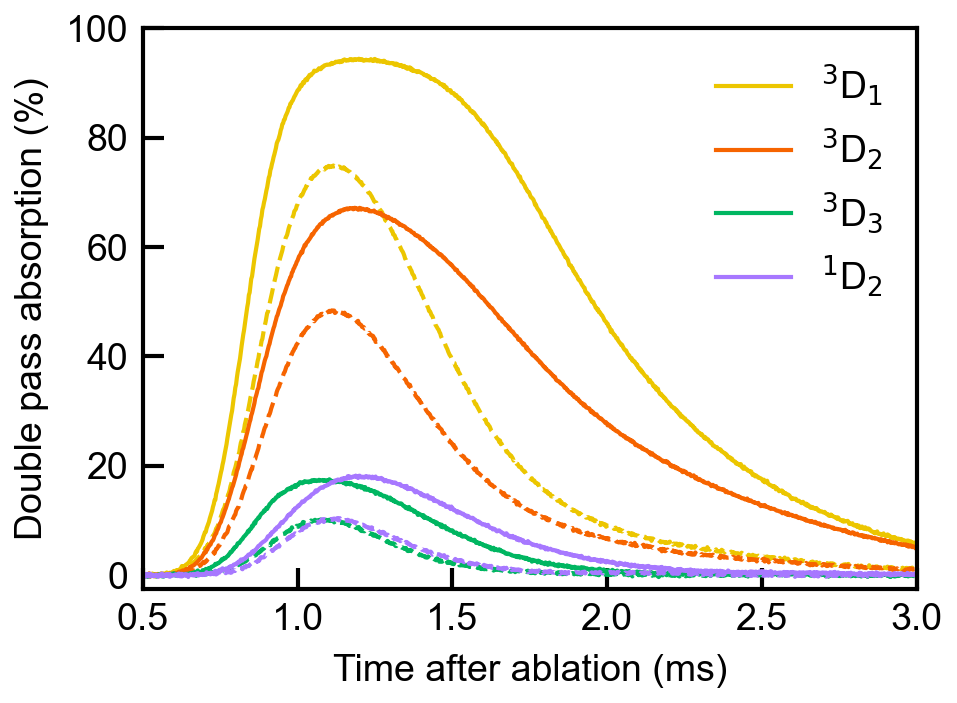}
    \caption{Absorption time of flight profiles showing the depletion of metastable D states of barium due to the presence of water vapour in the cell. Solid lines indicate time-of-flights without the reactant gas mixture sent in, and dashed lines indicate those with the reactant gas mixture sent in.}
    \label{fig:barium-depletion}
\end{figure}

To observe the contribution of the various metastable D states to the formation of BaOH, we probed the populations of the various $^3\mathrm D$ states in front of the cell aperture using the transitions shown in Figure~\ref{fig:baoh-spectrum-probe} (b). The time-of-flight profiles of these measurements are plotted in Figure \ref{fig:barium-depletion}. Precise lifetimes for the different states are unavailable apart from the $^3\mathrm P_1$ state referenced above. Therefore we rely on theoretical estimates for the following analysis. Significant population was observed in the $^3\mathrm D_1$ state despite being a weaker decay channel from the $^3\mathrm P_1$ state compared to that of the $^3\mathrm D_2$ state~\cite{Dzuba.Ginges.2006}. We expect redistribution over the metastable $^3\mathrm D$ states to arise due to collisional interactions between the barium atoms and the neon buffer gas inside the cell \cite{Ehrlacher.Huennekens.1994}. Additionally, a reduction of these populations was observed when we send in water, which in part is due to the quenching effects of a gas in the cell with vibrational degrees of freedom, but also due to the reaction of these barium atoms with the water molecules. A significant decrease in the all metastable D state populations is observed from which we conclude all states contribute in the formation of BaOH molecules. Distinguishing the quenching effects and the chemical reactions are beyond the scope of this paper. We also note the significant saturation effects leading to changes in shape of the time of flight in the absorption measurements. This prevents a detailed quantitative analysis of these processes.

\subsection*{Bubbler head pressure}
\begin{figure}[!ht]
    \centering
    \includegraphics[width=0.48\linewidth]{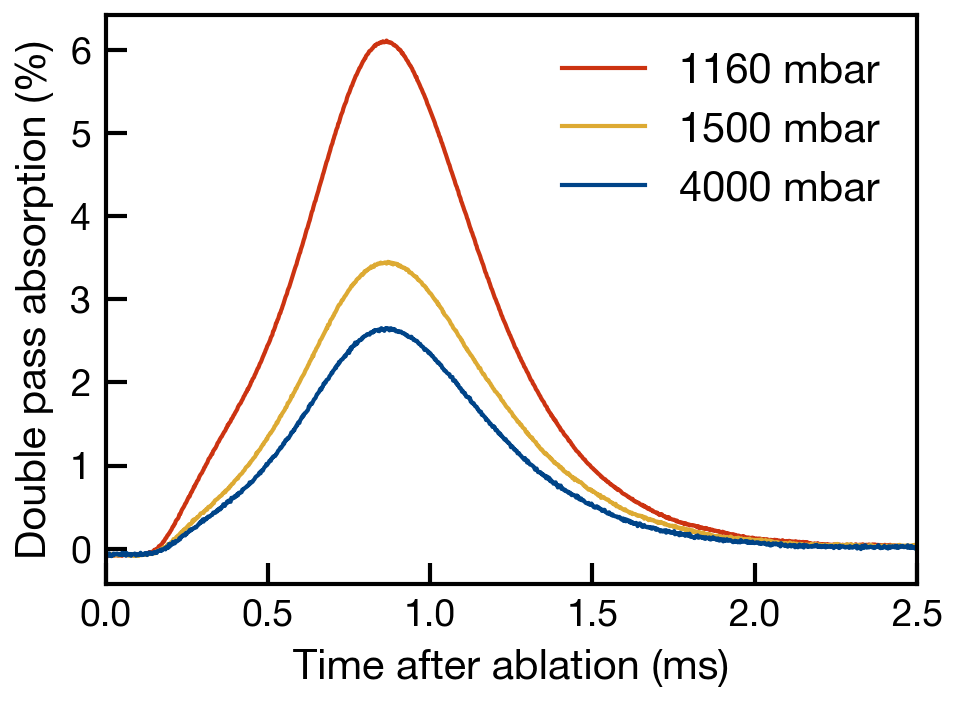}
    \caption{Molecular yield of BaOH as a function of bubbler head pressure.}
    \label{fig:bubblerpressure}
\end{figure}

\indent By controlling the head pressure of the bubbler the amount of water present in the carrier gas can be controlled. By decreasing the head pressure, more water is carried out of the bubbler for the same neon flow rate \cite{Betsch.1986}. The effect of the head pressure in the bubbler on the molecular yield can be seen in Figure \ref{fig:bubblerpressure}. It is important that the bubbler head pressure does not dip below the vapor pressure of water as this will cause the water in the bubbler to start boiling. This will lead to uncontrollable mixing ratios and result in unstable molecular yield. The pressure in the bubbler was set using a regulator, this limited fine control of the pressure. The operating range of this setup ranges from 1 to 4 bar abs. In future iterations this will be replaced by a digital pressure controller.

\subsection*{Effects of source parameters on the molecular beam}
We change various parameters of the source and observe the effects it
has on the molecular beam. We do this by characterizing the yield, velocity distribution, and rotational temperature of the molecular beam.

\begin{figure}[!ht]
    \centering
    \vspace{0.5cm}
    \includegraphics[width=0.9\linewidth]{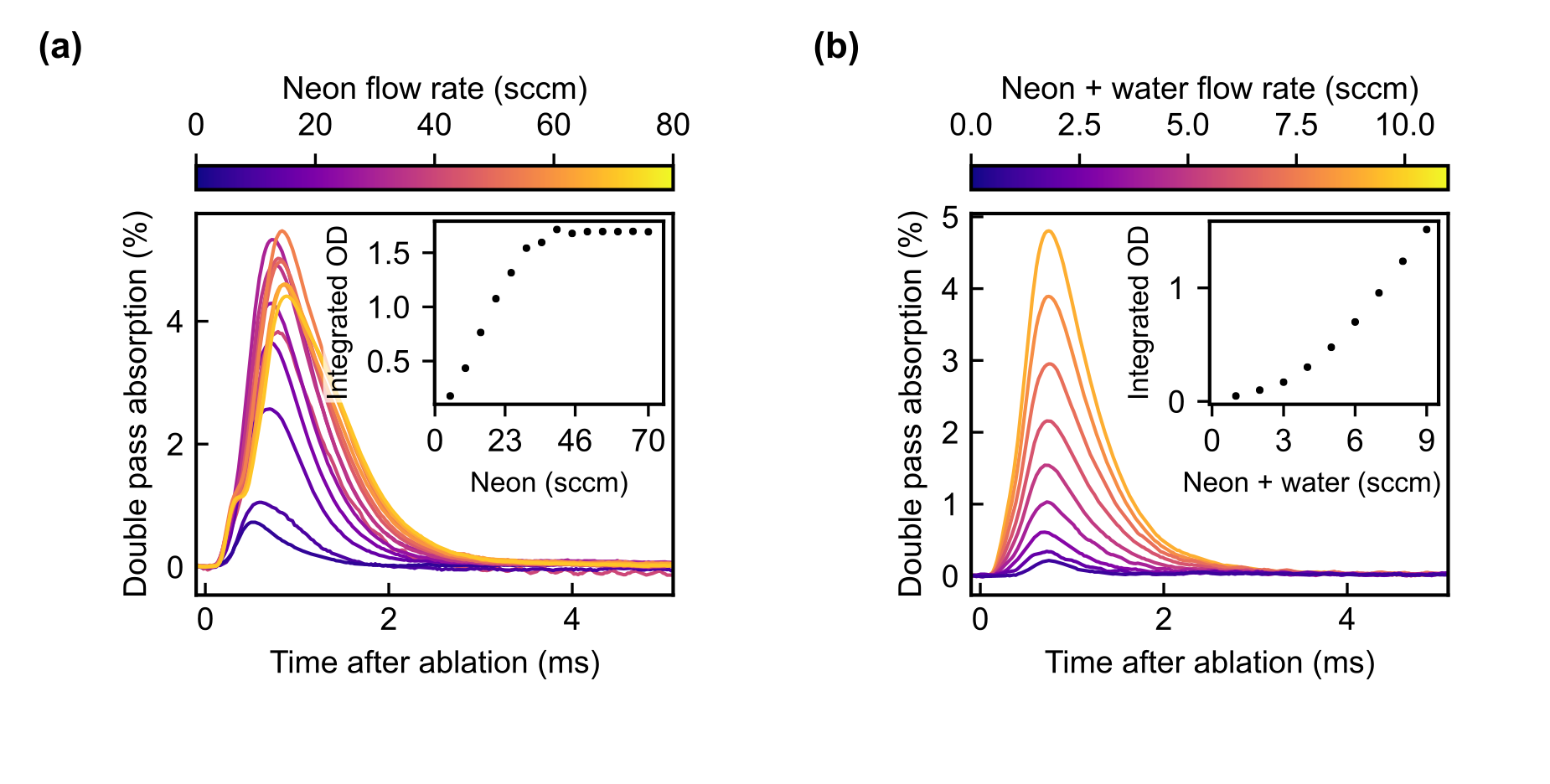}
    \caption{{\bf (a)}~Molecular yield of BaOH as a function of buffer gas flow rate, while the reactant mixture is sent in at 10~sccm. Inset shows the integrated optical density (OD) of the absorption signal.~{\bf (b)}~Molecular yield of BaOH as a function of reactant gas flow rate, when the buffer gas was sent in at 35~SCCM. Inset shows the integrated OD of the absorption signal.}
    \label{fig:flow-sweep}
\end{figure}

\subsection*{Molecular yield}
\indent We varied the flow rate of the neon buffer gas and the reactant gas mixture one at a time. When the neon flow rate was varied, the reactant flow rate was kept at 9~SCCM, and when the reactant flow rate was varied the neon flow rate was fixed at 35~SCCM. Doing so results in the time-of-flight profiles seen in Figure~\ref{fig:flow-sweep}. We note that the molecular yield appears to saturate around 30 SCCM of neon. At higher flow rates the peak of the absorption signal decreases but the width increases. Similarly, we observe increased yield with increased flow rates of the reactant gas, with no observable variations in beam properties. The maximum flow rate of the reactant flow is limited to 9~SCCM by the flow controller that was used. A higher molecular flux could likely be reached by using a higher flow through the bubbler, as can be seen from Figure~\ref{fig:flow-sweep} (b).

We are unable to directly determine the population of BaOH in the $N=1$ state since precise knowledge of the transition strength between these states is currently unavailable. Therefore, in order to produce an order of magnitude estimate, we assume that the transition strength is similar to an identical transition in BaF, for which the transition dipole moment is known. We use the $^SR_{21}(1)$ transition to estimate the population in the $N=1$ rotational level to be $\sim10^9$ per pulse under typical source conditions.

\subsection*{Forward velocity}
\begin{figure}[!ht]
    \centering
    \includegraphics[width=0.9\linewidth]{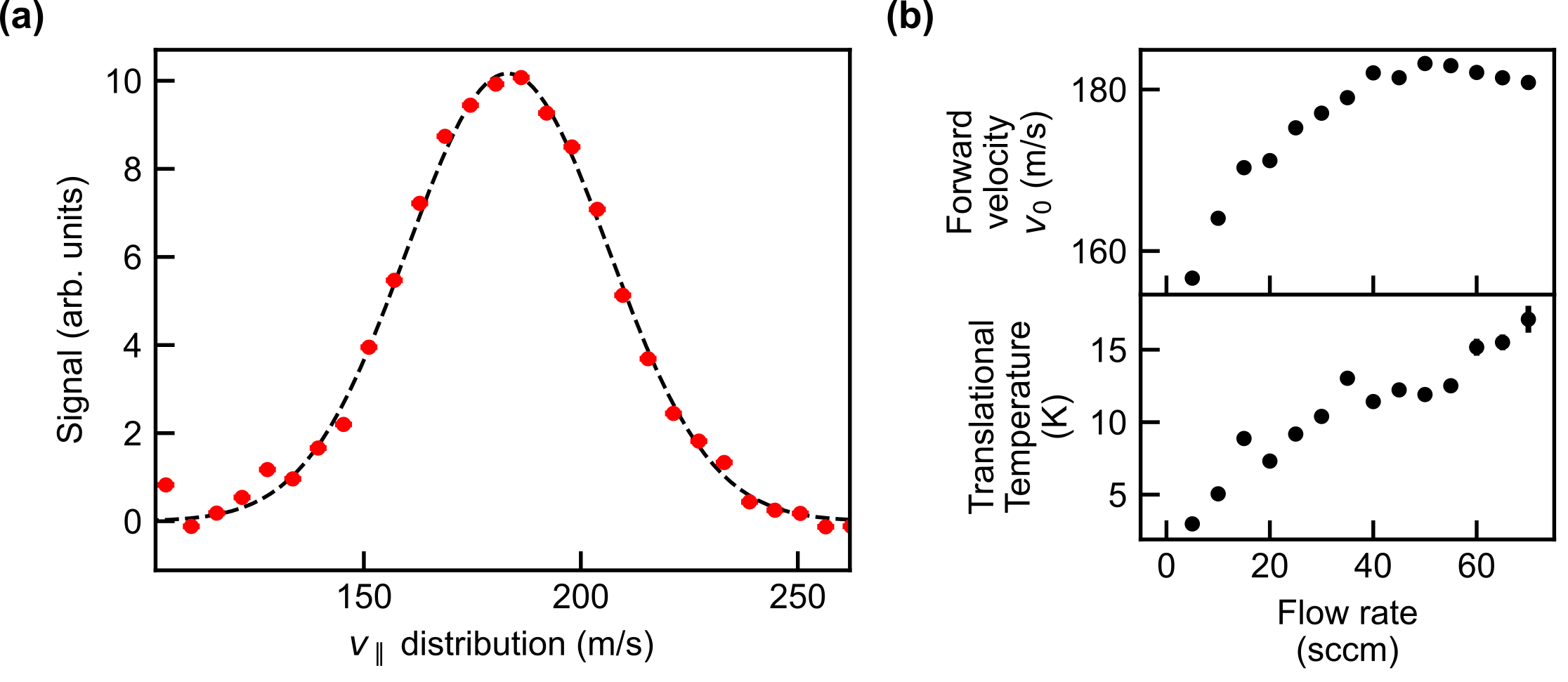}
    \caption{Forward velocity distribution of the BaOH beam. {\bf (a)} Forward velocity distribution of the molecules for buffer gas flow rate of 30~SCCM and reactant flow rate of 10~SCCM. The dashed line shows the fit to the velocity distribution function described in the text. {\bf (b)} Trends of forward velocity and translational temperature as a function of buffer gas flow rate derived from fitting to the aforementioned velocity distribution.}
    \label{fig:fit_results}
\end{figure}

The LIF detection zone lies 780~mm downstream from the source. Under the assumption that the trajectory is mostly in free-flight for this distance and that the pulse length is sufficiently short, we derive the velocity distribution of the molecules using the relation
$t = L/v_\parallel$. This approximation performs better for larger $L$. This provides us with the forward velocity distribution which we assume to be Maxwellian, allowing us to fit the distribution to the function, $f(v) = Av^2\exp(-M(v-v_0)^2/(2k_B T))$, where $A$ is the amplitude, $M$ is the mass of each molecule, $v_0$ corresponds to a forward velocity and $T$ is the translational temperature of the molecules~\cite{Truppe.Tarbutt.2018}. We perform this calculation for the data acquired from different buffer gas flow rates to characterize the forward velocity. The result is shown graphically in Figure~\ref{fig:fit_results} (a) when the buffer gas flow rate was set to 30 SCCM. From this fit, we plot the trends of the fitted forward velocity $v_0$ and the temperature $T$ as a function of the buffer gas flow rate in SCCM in Figure~\ref{fig:fit_results}.

Note that the forward velocity is observed to level off to about 180 m/s at around 35 -- 40 SCCM, similar to the trend observed for the integrated OD shown in Figure~\ref{fig:flow-sweep}. When the flow rate is increased further, not only does the molecular yield saturate and reach a stable forward velocity, resulting in a broadening of the arrival time distribution.

\subsection*{Rotational temperature}
\begin{figure}[!ht]
    \centering
    \vspace{0.5cm}
    \includegraphics[width=0.55\linewidth]{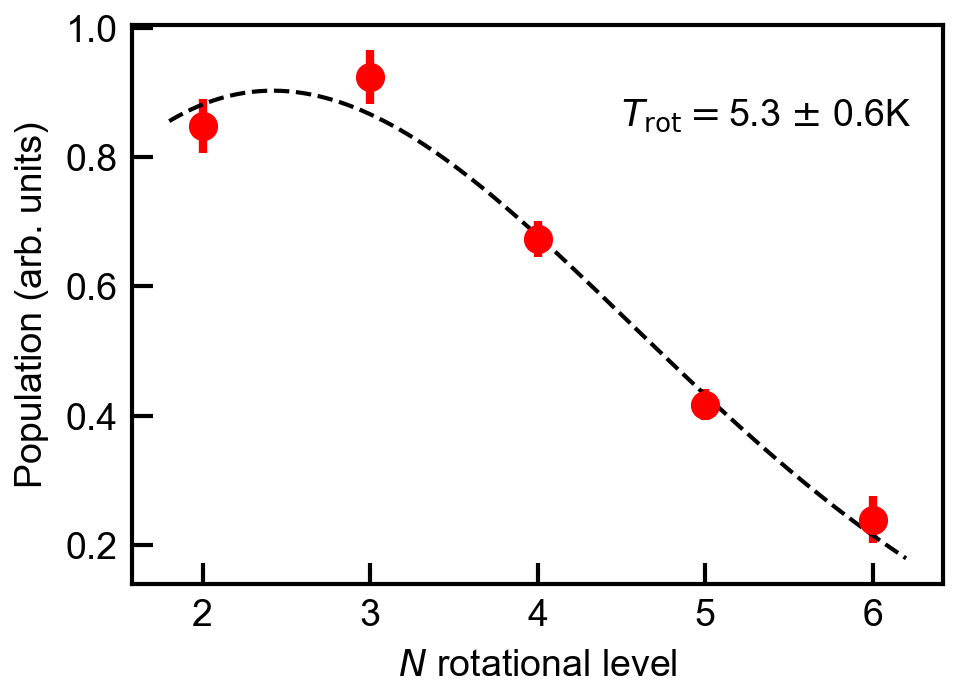}
    \caption{Population of the rotational levels in the $\tilde X^2\Sigma_{1/2}$ vibronic ground state. The peak on-resonance integrated OD was used as proxy for the population, which was then fit to the function $f(N) = A(2N+1)\exp[-BN(N+1)/T_\text{rot}]$ where $B = 0.3116$~K \cite{Fletcher.Ziurys.1995}.}
    \label{fig:rot-temp-fit}
\end{figure}

The populations of the rotational levels were determined by measuring the integrated optical densities of $^QQ(N'')$ transitions from $\tilde X ^2 \Sigma_{1/2}^+$ to $\tilde A ^2 \Pi_{3/2}$ for $2 \le N'' \le 6$~\cite{Kozyryev.Doyle.2015}. A Boltzmann distribution was fitted to the populations, from which a rotational temperature of $5.3\pm0.6$ K was extracted. This temperature was observed to change between 4 -- 6 K depending on source parameters and the time since the previous cleaning of the cell, which is known to affect the buffer gas cooling rate \cite{Mooij.Yin.2024}.

\subsection*{Bending mode population}

Optical transitions were observed between 12,057 -- 12,059~cm$^{-1}$. We believe that these are transitions from the bending ground state $\tilde X^2\Sigma(010)$ to the excited bending state $\tilde A^2\Pi(010)$. These will be discussed in a separate paper.

\section{Discussion}
\subsection*{Comparison of production methods and beam intensity}
We found that the molecular yield of BaOH achieved with both methods tested here is comparable, albeit slightly less, to that previously achieved for BaF molecules in the same source~\cite{Mooij.Yin.2024}. One advantage of using a salt target is the lack of the reactant gas that is needed to produce the molecules. However, the salt target was observed to disintegrate on a time scale of a week, which is unfeasible for molecular precision experiments.

These issues are not encountered with metal targets. Further investigation on the exact ingredients of the salt target -- especially in combination with direct laser heating -- may prove to be fruitful.

\subsection*{Bubbler}
Although a vapor seeded neon gas has the advantage of better control, it brings the complexity of having to heat the fill line used to inject the water into the cell. By controlling the head pressure in the bubbler, we have shown that it is possible to control the saturation level of the neon carrier gas. This allows the flow rate of the hot neon gas (283~K) to be lowered without decreasing the amount of water that reaches the cell, which reduces the overall heat and gas load on the cell. By lowering the bubbler pressure, a colder beam of BaOH molecules can be achieved without decreasing the molecular flux. The head pressure cannot be lowered too far as this will lead to condensation on the fill line walls, which leads to freezing or unstable molecular flux. Proper heating of the fill line is crucial to prevent this. The effect of the flow rate on the forward velocity of the molecules was minimal. Between 0.0 to 9.0~SCCM flows, the average forward velocity of the molecule remains around 180~m/s.

\subsection*{Enhancement}
The effect of the enhancement laser on the yield of the source is considerable. Integrated absorption signal over the duration of the pulse increases by a factor of 11 as shown in Figure~\ref{fig:enhancement} (a). The enhancement factor appears to be independent of many source parameters, such as the flow rates of the buffer and reactant gases, and other operational parameters such as the time since the cell was last cleaned. This is comparable to the increase in yield observed in the production of YbOH undergoing the same atomic transition \cite{Jadbabaie.Hutzler.2020}. By showing the dependence of the enhancement factor on the detuning of the excitation laser from the $^1\mathrm{S}_0 - ^3\mathrm{P}_1$ transition, we have demonstrated that the enhancement arises as a consequence of a resonant atomic effect.

The enhancement has the added advantage of increasing the relative isotope purity of the produced BaOH molecules. We primarily wish to produce $^{138}$BaOH, but $^{138}$Ba has a natural abundance of only 71.7\% \cite{Meija.Prohaska.2016}. By exciting the $^1\mathrm{S} - ^3\mathrm{P}$ transition specific to $^{138}$Ba, other barium isotopes remain in the ground \(^1\mathrm{S}\) state with reduced reactivity to water. This also provides an opportunity to selectively enhance the production of other BaOH isotopologues while not enhancing the production of the dominant isotopologue that may be of interest for future precision measurements on nuclear parity violating physics. We note that the degree of separation between the same transitions in the isotopologues will depend on their hyperfine structure.

\section{Conclusion}
In this paper, we have reported the production of a cryogenic buffer gas cooled beam of BaOH molecules. We have compared the production from salt and metal targets, used both methanol and water as reactant gases, and explored the enhancement of production of BaOH by exciting the electronic state of the barium atom. Once we settled on the optimal production scheme, we have characterised the forward velocity and intensity of the molecular beam as a function of the carrier gas flow rate and the reactant gas concentration.

We have found that using a metal precursor target in combination with a water-vapour seeded neon gas gives us the most consistent yield. By optically exciting barium atoms within the cell to a metastable state, we observe an enhancement in molecular yield by a factor of 11, achieving output levels comparable to those of BaF molecular beams. We note that this is an isotope-selective enhancement. The resulting beam exhibits rotational temperatures of $5.3\pm0.6$~K and forward velocities of 150 -- 180 m/s as expected, confirming that the molecular beam properties are similar to those of BaF beams produced in our laboratory \cite{Mooij.Yin.2024}. By controlling the pressure above the liquid in the bubbler, we can control the water-vapour content of the neon carrier gas. 

The cryogenic BaOH beam produced is a great starting point for precision measurements exploring fundamental physics. Next steps in the research are the characterization of the bending state, which is an excellent candidate for electron-EDM measurements~\cite{Bause.Hoekstra.2025}. In particular, further characterization of the transitions in the 12,057 -- 12,059 cm$^{-1}$ band are needed. In combination with dispersed laser-induced fluorescence (DLIF) spectra, this will enable us to further investigate efficient state population, optical cycling and laser cooling schemes.

\ack
We thank Rob Kortekaas and Leo Huisman for their expert technical assistance. We also thank Wander van der Meer and Kees Steinebach for insightful discussions and technical assistance. Lastly we thank Luk\'{a}\v{s} F. Pa\v{s}teka, I. Agust\'{\i}n Aucar and Anastasia Borschevsky for stimulating discussions. The work is financially supported by the Dutch Research Council (NWO) through projects VI.C.212.016, OCENW.M.21.098 and OCENW.XL21.XL21.074.

\section*{Bibliography}

\bibliographystyle{iopartnum.bst}

\begin{thebibliography}{10}
\bibitem{Roussy.Cornell.2023}
Roussy T~S, Caldwell L, Wright T, Cairncross W~B, Shagam Y, Ng K~B,
  Schlossberger N, Park S~Y, Wang A, Ye J and Cornell E~A 2023 {\em Science\/}
  {\bf 381} 46--50

\bibitem{Andreev.ACMECollaboration.2018}
Andreev V, Ang D~G, DeMille D, Doyle J~M, Gabrielse G, Haefner J, Hutzler N~R,
  Lasner Z, Meisenhelder C, O'Leary B~R, Panda C~D, West A~D, West E~P, Wu X
  and {ACME Collaboration} 2018 {\em Nature\/} {\bf 562} 355--360

\bibitem{Safronova.Clark.2018}
Safronova M~S, Budker D, DeMille D, Kimball D~F~J, Derevianko A and Clark C~W
  2018 {\em Rev. Mod. Phys.\/} {\bf 90}(2) 025008

\bibitem{Kozyryev.Doyle.2021}
Kozyryev I, Lasner Z and Doyle J~M 2021 {\em Phys. Rev. A\/} {\bf 103}(4)
  043313

\bibitem{Yu.Hutzler.2021}
Yu P and Hutzler N~R 2021 {\em Phys. Rev. Lett.\/} {\bf 126}(2) 023003

\bibitem{Wall.Carr.2015}
Wall M~L, Maeda K and Carr L~D 2015 {\em New Journal of Physics\/} {\bf 17}
  025001

\bibitem{Yu.Doyle.2019}
Yu P, Cheuk L~W, Kozyryev I and Doyle J~M 2019 {\em New Journal of Physics\/}
  {\bf 21} 093049

\bibitem{Albert.Preskill.2020}
Albert V~V, Covey J~P and Preskill J 2020 {\em Phys. Rev. X\/} {\bf 10}(3)
  031050

\bibitem{Hutzler.Hutzler.2020}
Hutzler N~R 2020 {\em Quantum Science and Technology\/} {\bf 5} 044011

\bibitem{Anderegg.Hutzler.2023}
Anderegg L, Vilas N~B, Hallas C, Robichaud P, Jadbabaie A, Doyle J~M and
  Hutzler N~R 2023 {\em Science\/} {\bf 382} 665--668

\bibitem{Augenbraun.Doyle.2023}
Augenbraun B~L, Anderegg L, Hallas C, Lasner Z~D, Vilas N~B and Doyle J~M 2023
  Chapter two - direct laser cooling of polyatomic molecules {\em Advances in
  Atomic, Molecular, and Optical Physics\/} ({\em Advances In Atomic,
  Molecular, and Optical Physics\/} vol~72) ed DiMauro L~F, Perrin H and Yelin
  S~F (Academic Press) pp 89--182

\bibitem{Bause.Hoekstra.2025}
Bause R, Balasubramanian N, Fikkers T, Prinsen E~H, Steinebach K, Jadbabaie A,
  Hutzler N~R, Aucar I~A, Pa\ifmmode~\check{s}\else \v{s}\fi{}teka L~F,
  Borschevsky A and Hoekstra S 2025 {\em Physical Review A\/} {\bf 111}(6)
  062815

\bibitem{Chamorro.Pasteka.2022}
Chamorro Y, Borschevsky A, Eliav E, Hutzler N~R, Hoekstra S and Pa{\v s}teka
  L~F 2022 {\em Physical Review A\/} {\bf 106} 052811

\bibitem{Takahashi.Miyamoto.2022}
Takahashi Y, Baba M, Enomoto K, Hiramoto A, Iwakuni K, Kuma S, Tobaru R and
  Miyamoto Y 2022 {\em The Astrophysical Journal\/} {\bf 936} 97

\bibitem{Kozyryev.2017}
Kozyryev I 2017 {\em Laser Cooling and Inelastic Collisions of the Polyatomic
  Radical SrOH\/} Ph.D. thesis Harvard University

\bibitem{Jadbabaie.Hutzler.2020}
Jadbabaie A, Pilgram N~H, K{\l}os J, Kotochigova S and Hutzler N~R 2020 {\em
  New Journal of Physics\/} {\bf 22} 022002

\bibitem{Isaev.Eliav.2017}
Isaev T~A, Zaitsevskii A~V and Eliav E 2017 {\em Journal of Physics B: Atomic,
  Molecular and Optical Physics\/} {\bf 50} 225101

\bibitem{Herschel.1823}
Herschel J~F~W 1823 {\em Transactions of the Royal Society of Edinburgh\/}
  445--460

\bibitem{James.Sugden.1955}
James C~G and Sugden T~M 1955 {\em Nature\/} {\bf 175} 333--334

\bibitem{Kinsey-Nielsen.Bernath.1986}
Kinsey-Nielsen S, Brazier C~R and Bernath P~F 1986 {\em The Journal of Chemical
  Physics\/} {\bf 84} 698--708

\bibitem{Fletcher.Ziurys.1995}
Fletcher D~A, Anderson M~A, Barclay W~L and Ziurys L~M 1995 {\em The Journal of
  Chemical Physics\/} {\bf 102} 4334--4339

\bibitem{Wang.Bernath.2008}
Wang J~G, Tandy J~D and Bernath P~F 2008 {\em Journal of Molecular
  Spectroscopy\/} {\bf 252} 31--36

\bibitem{Frey.Steimle.2011}
Frey S~E and Steimle T~C 2011 {\em Chemical Physics Letters\/} {\bf 512} 21--24

\bibitem{Hutzler.Doyle.2012}
Hutzler N~R, Lu H~I and Doyle J~M 2012 {\em Chemical Reviews\/} {\bf 112}
  4803--4827

\bibitem{Hutzler.Doyle.2011}
Hutzler N~R, Parsons M~F, Gurevich Y~V, Hess P~W, Petrik E, Spaun B, Vutha A~C,
  DeMille D, Gabrielse G and Doyle J~M 2011 {\em Physical Chemistry Chemical
  Physics\/} {\bf 13} 18976

\bibitem{Barry.DeMille.2011}
Barry J~F, Shuman E~S and DeMille D 2011 {\em Phys. Chem. Chem. Phys.\/} {\bf
  13}(42) 18936--18947

\bibitem{Wang.Zhao.2019}
Wang Q, Wang J, Chen Y and Zhao C 2019 {\em Solar Energy\/} {\bf 177} 99--107

\bibitem{Chichkov.Tunnermann.1996}
Chichkov B~N, Momma C, Nolte S, von Alvensleben F and T{\"u}nnermann A 1996
  {\em Applied Physics A\/} {\bf 63}(2) 109--115

\bibitem{Korner.Bergmann.1996}
K{\"o}rner C, Mayerhofer R, Hartmann M and Bergmann H~W 1996 {\em Applied
  Physics A\/} {\bf 63} 123--131

\bibitem{West.2017}
West E~P 2017 {\em A Thermochemical Cryogenic Buffer Gas Beam Source of ThO for
  Measuring the Electric Dipole Moment of the Electron\/} Ph.D. thesis Harvard
  University

\bibitem{Winnicki.Doyle.2024}
Winnicki A, Lasner Z~D and Doyle J~M 2024 {\em Phys. Rev. A\/} {\bf 110}(5)
  052804

\bibitem{Skoff.Hinds.2009}
Skoff S~M, Hendricks R~J, Sinclair C~D~J, Tarbutt M~R, Hudson J~J, Segal D~M,
  Sauer B~E and Hinds E~A 2009 {\em New Journal of Physics\/} {\bf 11} 123026

\bibitem{Tarbutt.Ezhov.2002}
Tarbutt M~R, Hudson J~J, Sauer B~E, Hinds E~A, Ryzhov V~A, Ryabov V~L and Ezhov
  V~F 2002 {\em Journal of Physics B: Atomic, Molecular and Optical Physics\/}
  {\bf 35} 5013

\bibitem{Aggarwal.Zapara.2018}
Aggarwal P, Bethlem H~L, Borschevsky A, Denis M, Esajas K, Haase P~A~B, Hao Y,
  Hoekstra S, Jungmann K, Meijknecht T~B, Mooij M~C, Timmermans R~G~E, Ubachs
  W, Willmann L and Zapara A 2018 {\em The European Physical Journal D\/} {\bf
  72} 197

\bibitem{Mooij.Yin.2024}
Mooij M~C, Bethlem H~L, Boeschoten A, Borschevsky A, Esajas K, Fikkers T~H,
  Hoekstra S, van Hofslot J~W~F, Jungmann K, Marshall V~R, Meijknecht T~B,
  Timmermans R~G~E, Touwen A, Ubachs W, Willmann L, Yin Y and eEDM
  collaboration N 2024 {\em New Journal of Physics\/} {\bf 26} 053009

\bibitem{Wright.Truppe.2023}
Wright S~C, Doppelbauer M, Hofs\"ass S, Schewe H~C, Sartakov B, Meijer G and
  and S~T 2022 {\em Molecular Physics\/} {\bf 121} e2146541

\bibitem{Ho.Liley.1972}
Ho C, Powell R~W and Liley P 1972 {\em J. Phys. Chem. Ref. Data\/} {\bf 1}(2)
  279--421

\bibitem{Zhang.Yang.2011}
Zhang Y, Evans J~R~G and Yang S 2011 {\em J. Chem. Eng. Data\/} {\bf 56}(2)
  328–--337

\bibitem{Brazier.Bernath.1986}
Brazier C~R, Ellingboe L~C, Kinsey-Nielsen S and Bernath P~F 1986 {\em Journal
  of the American Chemical Society\/} {\bf 108} 2126--2132

\bibitem{Scielzo.Potterveld.2006}
Scielzo N~D, Guest J~R, Schulte E~C, Ahmad I, Bailey K, Bowers D~L, Holt R~J,
  Lu Z~T, O’Connor T~P and Potterveld D~H 2006 {\em Physical Review A\/} {\bf
  73} 010501

\bibitem{Fernando.Bernath.1990}
Fernando W~T~M~L, Douay M and Bernath P~F 1990 {\em Journal of Molecular
  Spectroscopy\/} {\bf 144} 344--351

\bibitem{Davis.Mestdagh.1993}
Davis H~F, Suits A~G, Lee Y~T, Alcaraz C and Mestdagh J~M 1993 {\em The Journal
  of Chemical Physics\/} {\bf 98} 9595--9609

\bibitem{Mooij.Willmann.2024}
Mooij M~C, Bethlem H~L, Boeschoten A, Borschevsky A, Esajas K, Fikkers T~H,
  Hoekstra S, van Hofslot J~W~F, Jungmann K, Marshall V~R, Meijknecht T~B,
  Timmermans R~G~E, Touwen A, Ubachs W, Willmann L, Yin Y and eEDM
  collaboration N 2024 {\em New Journal of Physics\/} {\bf 26} 053009

\bibitem{Esajas.2021}
Esajas K 2021 {\em Intense slow beams of heavy molecules to test fundamental
  symmetries\/} Ph.D. thesis University of Groningen

\bibitem{Truppe.Tarbutt.2018}
Truppe S, Hambach M, Skoff S~M, Bulleid N~E, Bumby J~S, Hendricks R~J, Hinds
  E~A, Sauer B~E and Tarbutt M~R 2018 {\em Journal of Modern Optics\/} {\bf 65}
  648--656

\bibitem{Ehrlacher.Huennekens.1994}
Ehrlacher E and Huennekens J 1994 {\em Phys. Rev. A\/} {\bf 50} 4786--4793

\bibitem{Dzuba.Ginges.2006}
Dzuba V~A and Ginges J~S~M 2006 {\em Phys. Rev. A\/} {\bf 73}(3) 032503

\bibitem{Betsch.1986}
Betsch R~J 1986 {\em Journal of Crystal Growth\/} {\bf 77} 210--218

\bibitem{Kozyryev.Doyle.2015}
Kozyryev I, Baum L, Matsuda K, Olson P, Hemmerling B and Doyle J~M 2015 {\em
  New Journal of Physics\/} {\bf 17} 045003

\bibitem{Meija.Prohaska.2016}
Meija J, Coplen T~B, Berglund M, Brand W~A, Bi{\`e}vre P~D, Gr{\"o}ning M,
  Holden N~E, Irrgeher J, Loss R~D, Walczyk T and Prohaska T 2016 {\em Pure and
  Applied Chemistry\/} {\bf 88} 293--306
\end{thebibliography}

\end{document}